\documentclass[aps,twocolumn,prb,showpacs,preprintnumbers,amsmath,amssymb]{revtex4}
\newcommand{\lsim} 
 {\ \raise.35ex\hbox{$<$}\kern-0.75em\lower.5ex\hbox{$\sim$}\ }
\newcommand{\gsim}
 {\ \raise.35ex\hbox{$>$}\kern-0.75em\lower.5ex\hbox{$\sim$}\ }
\newcommand{\bras}[1]{\langle#1|}
\newcommand{\kets}[1]{|#1\rangle}

\newcommand{\means}[1]{\langle#1\rangle}

\def\journal #1#2#3#4{#1 {\bf #2}, #3 (#4)}
\def\PR{Phys.\ Rev.}

\def\PRB{Phys.\ Rev.\ B}
\def\PRX{Phys.\ Rev.\ X}
\def\PRL{Phys.\ Rev.\ Lett.}

\def\SSC{Solid State Commun.}

\def\JMMM{J.~Mag.~Mag.~Mat.}

\def\JPSJ{J.\ Phys.\ Soc.\ Jpn.}

\usepackage{graphicx}
\usepackage{dcolumn}
\usepackage{bm}
\usepackage{amsmath}
\usepackage{times}
\usepackage[usenames]{color}
\usepackage{natbib}

\usepackage{ulem}

\begin{document}
\title{Dynamical Jahn-Teller Effect in Spin-Orbital Coupled System}
\author{Joji~Nasu  and Sumio~Ishihara} 
 \affiliation{Department  of Physics,  Tohoku University,  Sendai 980-8578,  Japan} 
\date{\today}
\begin{abstract}  
Dynamical Jahn-Teller (DJT) effect in a spin-orbital coupled system on a honeycomb lattice is examined, motivated from recently observed spin-liquid behavior in Ba$_3$CuSb$_2$O$_9$. An effective vibronic Hamiltonian, where the superexchange interaction and the DJT effect are taken into account, is derived. We find that the DJT effect induces a spin-orbital resonant state where local spin-singlet states and parallel orbital configurations are entangled with each other. This spin-orbital resonant state is realized in between an orbital ordered state, where spin-singlet pairs are localized, and an antiferromagnetic ordered state. Based on the theoretical results, a possible scenario for Ba$_3$CuSb$_2$O$_9$ is proposed. 
\end{abstract}


\maketitle



%
%

%



\section{Introduction}\label{sec:intro}

No signs for long-range magnetic ordering down to the low temperatures, termed the quantum spin-liquid (QSL) state, are one of the fascinating states of matter in modern condensed matter physics.~\cite{Balents10} A number of efforts have been made to realize the QSL states theoretically and experimentally. One prototypical example is the well known one-dimensional antiferromagnets in which large quantum fluctuation destroys the long-range spin order even at zero temperature and realizes a spin-singlet state without any symmetry breakings. Another candidate for the QSL states has long been searched for in frustrated magnets. An organic salt $\kappa$-(BEDT-TTF)$_2$Cu$_2$(CN)$_3$ with a triangular lattice~\cite{Shimizu03} and an inorganic herbertsmithite ZnCu$_3$(OH)$_6$Cl$_2$ with a kagome lattice~\cite{Helton07} are some examples. Several theoretical scenarios for realization of the QSL, such as $Z_2$-spin liquid,~\cite{Read91} spin-nematic state,~\cite{Tsunetsugu06} spinon-deconfinement,~\cite{Senthil00} and so on, have been proposed so far. 

A transition-metal oxide of our present interest is a new candidate of the QSL state, Ba$_3$CuSb$_2$O$_9$, in which $S=1/2$ spins in Cu$^{2+}$ ions are responsible for the magnetism.~\cite{Zhou11,Nakatsuji12} There are no signs of magnetic orderings down to a few hundred mK, in spite of the effective exchange interactions of 30-50K. Temperature dependences of the magnetic susceptibility and the electronic specific heat are decomposed into the gapped component and the low-energy component; the latter is attributed to the so-called orphan spins.~\cite{Nakatsuji12} It was believed that the QSL behavior originates from the magnetic frustration in a Cu$^{2+}$ triangular lattice. Recent detailed crystal-structural analyses reveal that Cu ions are regularly replaced by Sb ions, and form a short-range honeycomb lattice.~\cite{Nakatsuji12} One characteristic in the present QSL system is that there is the orbital degree of freedom; two-fold orbital degeneracy in Cu$^{2+}$ is suggested by the three-fold rotational symmetry around a Cu$^{2+}$. Almost isotropic $g$-factors observed in the electron-spin resonance (ESR) provide a possibility of no static long-range orbital orders and novel roles of orbital on the QSL. 

In this paper, motivated from the recent experiments in Ba$_3$CuSb$_2$O$_9$, we examine a possibility of the QSL state in a honeycomb-lattice spin-orbital (SO) system. Beyond the previous theories for QSL in quantum magnets with the orbital degree of freedom,~\cite{Feiner97,Khaliullin97,Li98,Vernay2004,Lee12,Corboz12} the present study focuses on the dynamical Jahn-Teller (DJT) effect, which brings about a quantum tunneling between stable orbital-lattice states. This is feasible in the crystal lattice of Ba$_3$CuSb$_2$O$_9$, where O$_6$ octahedra surrounding Cu$^{2+}$ are separated from each other, unlike the perovskite lattice where nearest-neighboring (NN) two octahedra share an O$^{2-}$. The SO superexchange (SE) interactions between the separated NN Cu$^{2+}$ are comparable with the vibronic interactions, and a new state of matter is expected as a result of the cooperation between the on-site Jahn-Teller (JT) and inter-site SE interactions. We derive the low-energy electron-lattice model where the SE interaction, the JT effect, and the lattice dynamics are taken into account. It is discovered that a spin-orbital resonant state (SORS), where the two degrees of freedom are entangled with each other, is induced by the DJT effect. We examine connections of the quantum resonant state to the long-range ordered states, and provide a possible scenario for Ba$_3$CuSb$_2$O$_9$. 

In Sec.~\ref{sec:model}, a model Hamiltonian for a spin-orbital-lattice coupled system and calculation methods are introduced. In Sec~\ref{sec:result}, numerical results are presented. Section~\ref{sec:discussion-summary} is devoted to discussion and summary.

\section{Model and method}
\label{sec:model}

\begin{figure}
\begin{center}
\includegraphics[width=\linewidth]{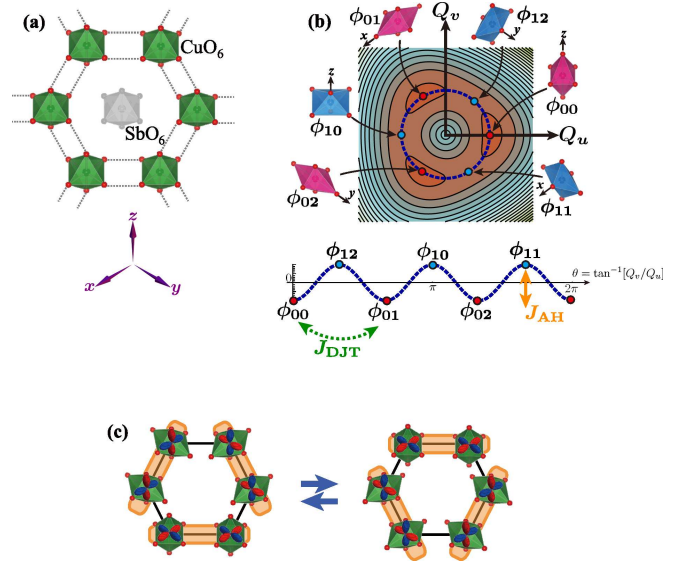} 
\caption{(color online). (a)~A honeycomb lattice structure for Ba$_3$CuSb$_2$O$_9$.~\cite{coord} (b)~A contour map of the lower AP surface on the $Q_u$-$Q_v$ plane. A blue broken line shows a circle for $\rho=\rho_0$. The lower panel represents the AP along this circle as a function of $\theta=\tan^{-1}[Q_v/Q_u]$. Schematic O$_6$ distortions at the potential minima and maxima are also shown. (c)~A schematic picture for the SORS. Shaded bonds represent the spin-singlet and parallel-orbital bonds. 
}
\label{lattice}
\end{center}
\end{figure}

First we set up the model which consists of the SE interactions between the Cu ions in a honeycomb lattice, and the local vibronic coupling between the Cu $d$ orbitals and O$_6$ octahedron. The Hamiltonian is given by ${\cal H}= {\cal H}_{\rm exch}+{\cal H}_{\rm JT}$. In the first term for the SE interactions, the doubly-degenerate $3d_{3z^2-r^2}$ and $3d_{x^2-y^2}$ orbitals are introduced at each site. The SE interactions are derived from the extended $dp$-type Hamiltonian where the $3d$ orbitals for a Cu ion and $2p$ orbitals for a O ion are introduced, and the on-site electron-electron interactions and the Cu-O and O-O electron transfers are considered. All possible exchange paths between the NN Cu pairs are taken into account. Details are given in the Supplemental Material (SM).~\cite{suppl} The obtained Kugel-Khomskii type SO coupled Hamiltonian is gven by
\begin{align}
{\cal H}_{\rm exch}&=J_{\rm SE}\sum_{\langle ij \rangle_l}
\Bigl[\bm{S}_i\cdot\bm{S}_j+\bar{J}_{\tau\tau}\tau_i^l \tau_j^l+\bar{J}_{\bar{\tau}\bar{\tau}}\bar{\tau}_i^l \bar{\tau}_j^l+\bar{J}_{yy}T_i^y T_j^y\nonumber\\
&+\bar{J}_{ss\tau}\bm{S}_i\cdot\bm{S}_j (\tau_i^l +\tau_j^l)
+\bar{J}_{ss\tau\tau}\bm{S}_i\cdot\bm{S}_j \tau_i^l \tau_j^l\nonumber\\
&+\bar{J}_{ss\bar{\tau}\bar{\tau}}\bm{S}_i\cdot\bm{S}_j \bar{\tau}_i^l \bar{\tau}_j^l+\bar{J}_{ssyy}\bm{S}_i\cdot\bm{S}_j T_i^y T_j^y\Bigr], 
\label{eq:3}
\end{align}
where NN $ij$ sites along a direction $l(=x,y,z)$ [see Fig.~\ref{lattice}(a)~\cite{coord}] is denoted by $\langle ij \rangle_l$. We introduce the spin operator ${\bm S}_i$ and the pseudo-spin (PS) operator ${\bm T}_i$ for the orbital degree of freedom with amplitudes of 1/2. The eigenstate for $T^z=+1/2$ ($-1/2$) describes a state where the $d_{3z^2-r^2}$ ($d_{x^2-y^2}$) orbital is occupied by a hole. For convenience, we introduce the bond-dependent PS operators defined by 
\begin{align}
 \tau_i^l=\cos \left( \frac{2 \pi n_l}{3}\right) T_i^z -\sin \left (\frac{2 \pi n_l}{3} \right) T_i^x
\end{align}
and
\begin{align}
 \bar{\tau}_i^l=\cos \left(\frac{2 \pi n_l}{3} \right) T_i^x +\sin \left( \frac{2 \pi n_l}{3} \right) T_i^z,
\end{align}
where $(n_z,n_x, n_y)=(0,1,2)$. The exchange constants are given by the energy parameters in the $dp$-type Hamiltonian,~\cite{suppl,Slater} and are normalized by the representative SE interaction $J_{\rm SE}$, a coefficient of $\bm{S}_i\cdot\bm{S}_j$. 
It is shown that ${\cal H}_{\rm exch}$ in NN two sites favors an antiferromagnetic (AFM) and ferro-type orbital configuration in a wide parameter region which includes a parameter set for Ba$_3$CuSb$_2$O$_9$.~\cite{Mizuno98,suppl} This is in contrast to the results of the Kugel-Khomskii type SO Hamiltonian in a square lattice, where a ferromagnetic and antiferro-type orbital configuration is realized.
Among a number of terms in the Hamiltonian, the $\bm{S}_i \cdot \bm{S}_i$, $\tau_i \tau_j$, $\bm{S}_i \cdot \bm{S}_j \tau_i \tau_j$, and $\bar{\tau}_i \bar{\tau}_j$ terms are essential for the SORS of our main interest.

The second term of the Hamiltonian, ${\cal H}_{\rm JT}$, describes the local vibronic coupling in each CuO$_6$ octahedra. We consider the $e \times E$ JT problem where the degenerate $d_{3z^2-r^2}$ and $d_{x^2-y^2}$ orbitals are coupled with the $E$-symmetric O$_6$ vibrations, denoted by $Q_u$ and $Q_v$. The harmonic vibration, the linear JT coupling and the anharmonic lattice potential, are taken into account. We focus on the low-energy vibronic mode, i.e. the rotational motion in the $Q_u$-$Q_v$ plane along the bottom in the lower adiabatic potential (AP) plane [see Fig.~\ref{lattice}(b)]. The Hamiltonian is a well-known form given by~\cite{Koehler40,OBrien64,Slonczewski69,Williams69,Englman_text,Bersuker_text}
\begin{align}
 {\cal H}_{\rm rot}=-\frac{1}{2M \rho_0^2} \frac{\partial^2}{\partial \theta^2} +B \rho_0^3 \cos 3\theta , 
\end{align}
with an oxygen mass $M$, an amplitude of the distortion $\rho_0$, an angle $\theta=\tan^{-1}(Q_v/Q_u)$ in the AP, and an anharmonic potential $B$. This model represents the angle motion under the three-fold potential, which takes minima (maxima) at $\theta_{0 \nu}=2\pi \nu /3$ ($\theta_{1 \nu}=\pi+2\pi \nu /3$) with $\nu=(0,1,2)$. These angles correspond to the cigar-type $[(3z^2-r^2)$-type] and the leaf-type [$(x^2-y^2)$-type] lattice distortions, respectively, as shown in Fig.~\ref{lattice}(b). The low-energy states in ${\cal H}_{\rm rot}$ are well described by the six Wannier-type vibronic functions,~\cite{Halperin69} $\kets{\Phi_{\mu\nu}}$, localized around $\theta_{\mu \nu}$, as shown schematically in the lower panel of Fig.~\ref{lattice}(b). Then, the low-energy vibronic Hamiltonian is given by
\begin{align}
{\cal H}_{\rm JT}=
  \sum_{i \mu=(0,1)}\frac{\sigma_\mu}{2}\Bigl[-J_{\rm AH}\sum_{\nu}\kets{\Phi_{i \mu\nu}}\bras{\Phi_{i \mu\nu}}\nonumber\\
+J_{\rm DJT}\sum_{\nu\neq \nu' }\kets{\Phi_{i \mu\nu}}\bras{\Phi_{i \mu\nu'}}\Bigr],
\label{eq:djt}
\end{align}
where $(\sigma_0, \sigma_1)=(1,-1)$. The first and second terms describe the potential in the angle space and the tunneling motions, respectively. The energy constants, $J_{\rm AH}$ and $J_{\rm DJT}$, are positive and are of the order of $B\rho_0^3$, and $1/(M\rho_0^2)$, respectively. A condition $J_{\rm DJT}/J_{\rm AH}< 1/2$ is required to reproduce the original energy levels, but we regard $J_{\rm DJT}/J_{\rm AH}(\equiv j_D)$ as a free parameter, for convenience. The lowest energy state is a doublet, corresponding to the clockwise and counter-clockwise rotations in the $\theta$ space. The so-called vibronic reduction factor proposed by Ham,~\cite{Ham65,Ham68} i.e, a reduction of the PS moment due to the DJT effect is 1/2. The detailed derivation of ${\cal H}_{\rm JT}$ is given in SM.~\cite{suppl}

There are three principal energy parameters in the Hamiltonian, the SE interaction $J_{\rm SE}$, the anharmonic potential $J_{\rm AH}$ and the DJT effect $J_{\rm DJT}$. The magnitude of $J_{\rm SE}$ is about 1-10meV, which is smaller than the exchange interaction in the high-$T_c$ cuprates because of a large distance between the NN Cu sites. Both the energy scales of $J_{\rm AH}$ and $J_{\rm DJT}$ are 1-30meV.~\cite{Fernandez10,Abtew11} Since the three parameters are in the same order of magnitude, competitions and cooperation among them are realized.

The Hamiltonian is analyzed by the exact-diagonalization (ED) method combined with the mean-field (MF) approximation, and the quantum Monte-Carlo simulation (QMC)~\cite{ALPS,Todo} with the MF approximation, termed the ED+MF and the QMC+MF methods, respectively. We introduce mainly the results by the ED+MF method. The MF type decouplings are introduced in the exchange interactions which act on the edge sites of clusters. The Hamiltonian for a 6-site cluster under the MFs is diagonalized by the Lanczos algorithm, and the MFs are determined consistently with the states inside of the cluster. This method is equivalent to the hierarchical mean-field method~\cite{Isaev2009,nasu11} where the no long-range ordered phase obtained by the large cluster size~\cite{Jiang2012,Lauchli05} is reproduced.  The adopted parameter values are $J_{\rm SE}/J_{\rm AH}=0.15$ and are given in Ref.~\onlinecite{suppl}. Amplitude of the DJT effect, i.e., $j_D=J_{\rm DJT}/J_{\rm AH}$, is varied. We find that the obtained results do not depend qualitatively on the parameter $J_{\rm SE}/J_{\rm AH}$ between 0.075 and 1. The exchange Hamiltonian ${\cal H}_{\rm exch}$ is also analyzed by the MF approximation and by the ED method as supplementary calculations. 

\section{Result}\label{sec:result}

\begin{figure}
\begin{center}
\includegraphics[width=\linewidth]{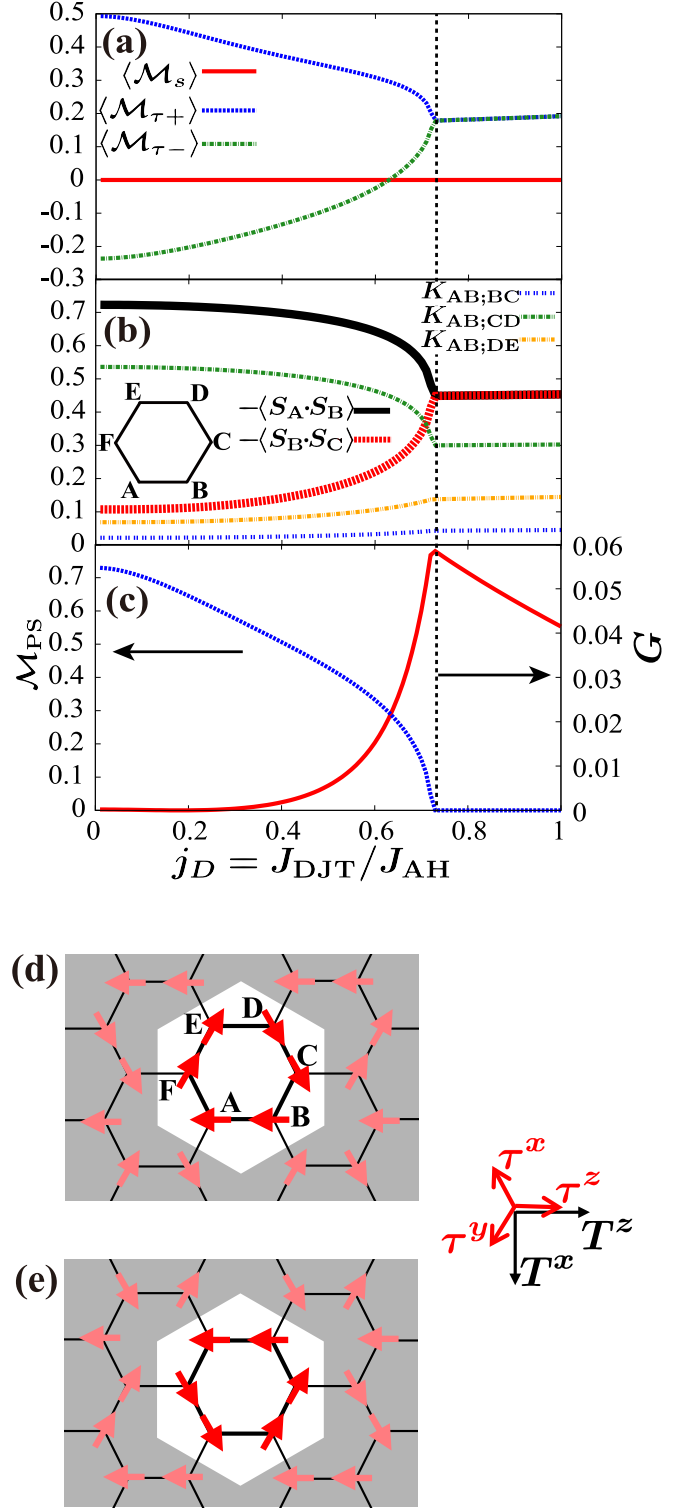} 
\caption{
(color online). (a)~Staggered spin moment $\mathcal{M}_s$, and two PS moments $\mathcal{M}_{\tau+}$ and $\mathcal{M}_{\tau-}$.  (b)~Spin-correlations $\means{\bm{S}_{\rm{A}}\cdot\bm{S}_{\rm{B}}}$ and $\means{\bm{S}_{\rm{B}}\cdot\bm{S}_{\rm{C}}}$ (bold lines), and spin-dimer correlations $K_{\rm AB;BC}$, $K_{\rm AB;CD}$ and $K_{\rm AB;DE}$ (thin lines). The inset shows a 6-site cluster. (c) SO correlation function $G$, and the three-fold orbital order parameter $\mathcal{M}_{\rm PS}$. The broken line indicates $j_{Dc}$. (d)~(e)~Two kinds of the three-fold orbital ordered states. 
}
\label{bethe}
\end{center}
\end{figure}

Spin and orbital structures are monitored by the staggered spin moment given by
\begin{align}
 \mathcal{M}_s=\frac{1}{6}\sum_{i}(-1)^i S_i^z , 
\end{align}
and the two PS moments defined by
\begin{align}
 \mathcal{M}_{\tau+}=-\frac{1}{6}
 \left (\tau_{{\rm A}}^z+\tau_{{\rm B}}^z+\tau_{{\rm C}}^x+\tau_{{\rm D}}^x+\tau_{{\rm E}}^y+\tau_{{\rm F}}^y \right) , 
\end{align}
and
\begin{align}
 \mathcal{M}_{\tau-}=-\frac{1}{6}
 \left(\tau_{{\rm A}}^x+\tau_{{\rm B}}^y+\tau_{{\rm C}}^y+\tau_{{\rm D}}^z+\tau_{{\rm E}}^z+\tau_{{\rm F}}^x \right),
\end{align}
where subscripts A-F indicate sites in the cluster [see the inset of Fig.~\ref{bethe}(b)]. We note that $\mathcal{M}_{\tau+}$ and $\mathcal{M}_{\tau-}$ take their maxima of 0.5 in the  three-fold orbital ordered states shown in Figs.~\ref{bethe}(d) and (e), respectively. A difference between the two, $\mathcal{M}_{\rm PS} \equiv \langle \mathcal{M}_{\tau+} - {\mathcal{M}}_{\tau-} \rangle$, is regarded as an amplitude of the symmetry breaking. The numerical results are plotted in Fig.~\ref{bethe}(a). There is a critical value of $j_D$, termed $j_{Dc}(=0.75)$. For $j_D \ll j_{Dc}$, $\mathcal{M}_{\tau+} \sim 0.5$ and $\mathcal {M}_{\tau-}\sim -0.25$, implying a symmetry breaking due to the three-fold orbital order shown in Fig.~\ref{bethe}(d). This orbital order is also suggested by the analyses of ${\cal H}_{\rm exch}$. With increasing $j_D$, absolute values of ${\cal M}_+$ and ${\cal M}_-$ are reduced. These reductions are reproduced by the QMC+MF method. Above $j_{Dc}$, $\langle \mathcal{M}_{\tau+} \rangle=\langle \mathcal{M}_{\tau-} \rangle$, interpreted as a superposition of the two PS configurations. 

As for the spin sector, neither a finite staggered moment [see Fig.~\ref{bethe}(a)], nor a finite local moment $\langle S_i^z \rangle$ at each site, are obtained in a whole parameter region of $j_D$. As shown in Fig.~\ref{bethe}(b), there are two inequivalent NN spin correlations $\means{\bm{S}_i\cdot\bm{S}_j}$ for $j_D<j_{Dc}$: large values are shown in the bonds where the PSs are parallel with each other. On the other hand, for $j_D > j_{Dc}$, $\means{\bm{S}_i\cdot\bm{S}_j}$ for all bonds are equivalent. Spin-dimer correlations defined by $K_{ij;kl}=\langle (\bm{S}_i\cdot\bm{S}_j)(\bm{S}_k\cdot\bm{S}_l) \rangle$ are also plotted in Fig.~\ref{bethe}(b) where $K_{\rm AB;BC}$, $K_{\rm AB;CD}$ and $K_{\rm AB;DE}$ are the bond-correlation functions for the NN bonds, the 2nd-NN bonds, and the 3rd-NN bonds, respectively. The 2nd-NN bond correlation function $K_{\rm AB;CD}$ is the largest in a whole parameter region, and $K_{\rm AB;CD}$ in $j_D<j_{Dc}$ is larger than that in $j_D>j_{Dc}$. The results in $j_D<j_{Dc}$ are interpreted as a valence-bond solid state, where spin-singlet pairs are localized at the bonds, in which the NN PSs are parallel with each other. This SO structure is also confirmed in the analysis by the QMC+MF method. Above $j_{Dc}$, this classically localized PS state associated with the localized single pairs is changed into a quantum superposition of the PS configurations. The spin-singlet dimers are no longer localized in specific bonds, suggested by a reduction of $K_{\rm AB;CD}$ and enhancements of $K_{\rm AB;BC}$ and $K_{\rm AB;DE}$.  

\begin{figure}
\begin{center}
\includegraphics[width=\linewidth]{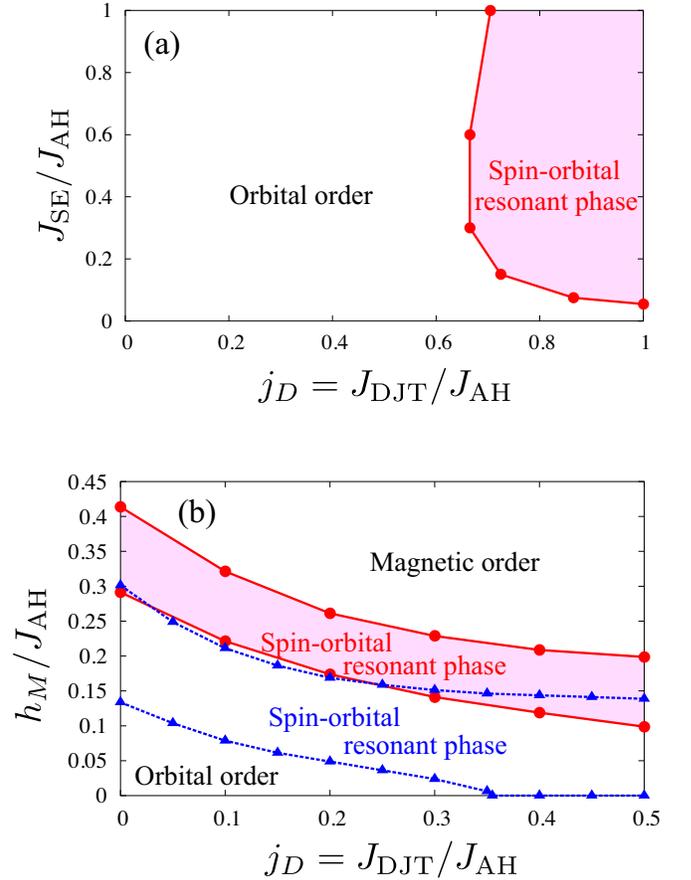} 
\caption{(color online). 
(a)~Phase diagram on the plane of $J_{\rm DJT}/J_{\rm AH}$ and $J_{\rm SE}/J_{\rm AH}$. (b)~Phase diagram on the plane of $J_{\rm DJT}/J_{\rm AH}$ and $h_{M}/J_{\rm AH}$. The circles and triangles represent the phase boundaries with and without the interaction between the octahedra, respectively. The parameter of this interaction is chosen to be $K/J_{\rm AH}=0.1$ (see the details in the SM~\cite{suppl}). 
} 
\label{phase2}
\end{center}
\end{figure}

We expect from these data that, above $j_{Dc}$, the local spin-singlet and the parallel PS configuration are strongly entangled with each other. This is directly confirmed by the SO correlation function defined by~\cite{Oles2006}
\begin{align}
 G=\Bigl[\frac{1}{6} \sum_{\means{ij}_l}G_{ij}^l\Bigr]^2 , 
\end{align}
with
\begin{align}
 G_{ij}^l=16
 \left[\means{(\bm{S}_i\cdot\bm{S}_j) (\tau_i^l \tau_j^l)}-\means{\bm{S}_i\cdot\bm{S}_j}\means{\tau_i^l \tau_j^l} \right].
\end{align}
 The results are presented in Fig.~\ref{bethe}(c). Spin and orbital sectors are decoupled at $j_D=0$, and are strongly entangled near and above $j_{Dc}$. This is consistent with the picture where the spin-singlet and the parallel PS configuration are realized as a quantum mechanical superposition.~\cite{Feiner97,Li98,Oles2006} The phase diagram on a plane of $J_{\rm DJT}$-$J_{\rm SE}$ is shown in Fig.~\ref{phase2}(a). The SORS diminishes both in the weak and strong $J_{\rm SE}$ limits. That is, the SORS is realized by interplay of $J_{\rm SE}$ and $J_{\rm DJT}$.

To examine a connection of the present SORS to the long-range spin ordered state in the honeycomb-lattice Heisenberg model, we release the orbital degeneracy by applying the artificial external field on the orbital-lattice sector. The is given by
\begin{align}
 {\cal H}_M=-h_M \sum_{i \mu \nu \nu'} \sigma_\mu F_{\nu \nu'} |  \Phi_{i \mu \nu} \rangle \langle  \Phi_{i \mu \nu'}| , 
\end{align}
where $F_{\nu \nu'}=\frac{i}{\sqrt{3}}\sum_l\varepsilon_{l\nu \nu'}$ with the Levi-Civita completely-antisymmetric tensor $\varepsilon_{l\nu\nu'}$.~\cite{suppl} The artificial field makes clockwise and counter-clockwise rotations in the $\theta$ space inequivalent, and lift the ground state degeneracy. Then, the Hamiltonian is reduced into the AFM Heisenberg model without the orbital degree of freedom. The phase diagram on a plane of $j_D$ and $h_{M}$ is shown in Fig.~\ref{phase2}(b). It is obtained that the N\'eel order appears for large $h_M$. The SORS is realized in between the spin ordered state and the orbital ordered state which is realized in small $j_D$ and $h_M$. 

So far, each O$_6$ lattice vibration is assumed to be independent from each other. Here we show that the SORS is realized in more realistic parameter space, when the crystal lattice effect is taken into account. The interactions ${\cal H}_{\rm inter}$ between the NN $\rm O_6$ octahedra are modeled by introducing the spring constant between the NN octahedra $K$ (see details in Ref.~\onlinecite{suppl}). As shown in Fig.~\ref{phase2}(b), the SORS is shifted to the lower side of $j_D$ and $h_M$. Without the artificial field ($h_M=0$), $j_{Dc}$ is decreased down to 0.35. This result satisfies the condition of $j_D<0.5$, in which ${\cal H}_{\rm JT}$ is valid as an effective Hamiltonian for the low-energy vibronic states of ${\cal H}_{\rm rot}$.

\begin{figure}
\begin{center}
\includegraphics[width=\linewidth]{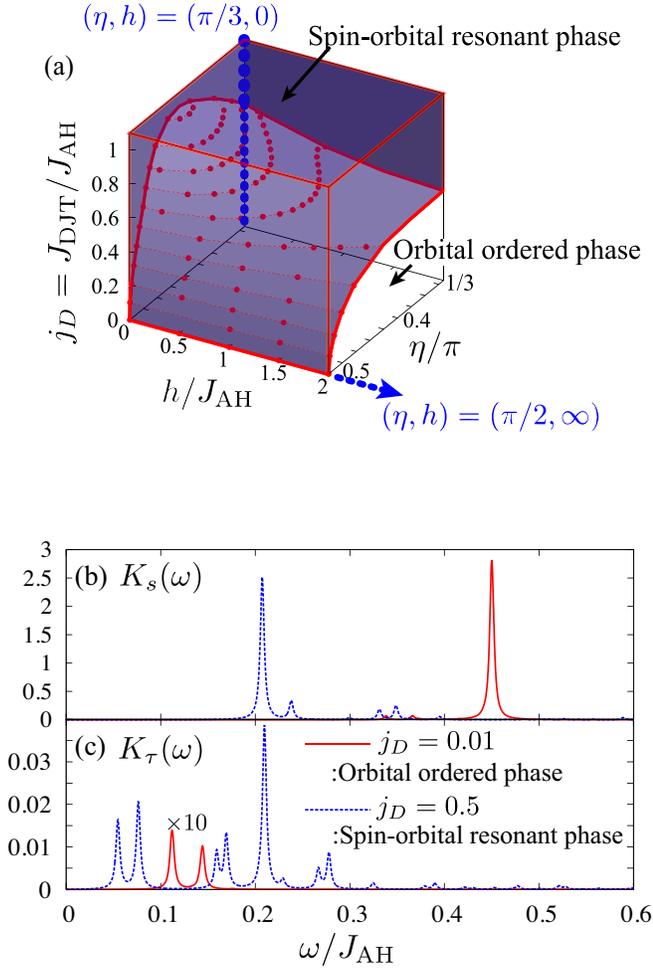} 
\caption{(color online). (a)~Phase diagram for the model where the generalization of the electron transfer, $\eta$, and the staggered magnetic field, $h$, are taken into account [see text]. The SE interaction model $\mathcal{H}_{\rm exch}$ and the orbital-only model $\mathcal{H}_{\rm orb}$ correspond to $(\eta, h)=(\pi/3, 0)$ and $(\pi/2, \infty)$, respectively. A shaded area shows the SO resonant phase. 
(b) The dynamical spin-correlation function $K_s(\omega)$, and (c) the dynamical orbital-correlation function $K_\tau(\omega)$ in the orbital-ordered phase ($j_D=0.01$) and in the SO resonant phase ($j_D=0.5$).
Parameter values are chosen to be the same with those in Fig.~\ref{phase2}(b).
} 
\label{eta_dynamics}
\end{center}
\end{figure}

We have shown that the present SORS emerges under the quantum orbital state. A similar orbital state is known in the honeycomb lattice ``orbital-only'' model without the spin degree of freedom, given by
\begin{align}
 {\cal H}_{\rm orb}=J\sum_{\langle ij \rangle_l} \tau_i^l \tau_j^l.
\end{align}
 Instead of a conventional long-range order, a quantum superposition of the orbital PSs is realized in ${\cal H}_{\rm orb}$.~\cite{Nasu08} Here, we connect the present SE Hamiltonian ${\cal H}_{\rm exch}$ in Eq.~(\ref{eq:3}) to ${\cal H}_{\rm orb}$, and examine a relation between the two orbital states. We generalize the electron transfer as $t_{pd} \rightarrow t_{pd}(\eta)$ by introducing a parameter $\eta$, and apply the staggered magnetic field as
\begin{align}
 {\cal H}_h=-h\sum_i(-1)^i S_i^z.
\end{align}
 Detailed procedures are explained in the SM.~\cite{suppl} In the $(\eta, h)$ parameter space, the present model and the orbital-only model are located at $(\eta, h)=(\pi/3, 0)$ and $(\pi/2, \infty)$, respectively. In Fig.~\ref{eta_dynamics}(a), the phase diagram as functions of $h$, $\eta$ and $j_D$ is presented. The SORS at $(\eta, h)=(\pi/3, 0)$ and $j_{D}>j_{Dc}=0.75$  continuously connects to the orbital resonant state in the orbital-only model realized in $(\eta, h)=(\pi/2, \gg J_{\rm AH})$. This result implies that the present SORS belongs to the same class of the orbital resonance state in the orbital-only model, and supports the physical picture given in Fig.~\ref{lattice}(c). It is worth noting that the orbital resonant state in ${\cal H}_{\rm orb}$ remains in the infinite size limit.~\cite{Nasu08} We suppose that the present SORS is survived even in large cluster systems.

\section{Discussion and Summary}\label{sec:discussion-summary}

Based on the calculations, we propose a scenario for Ba$_3$CuSb$_2$O$_9$.~\cite{Zhou11,Nakatsuji12} The x-ray diffraction experiments suggest a short-range honeycomb-lattice domain of the order of 10\AA, which justifies the present finite-size cluster analyses to mimic the realistic situation. The observed positive Weiss constant is not trivial in conventional orbital degenerate magnets where the ferromagnetic interaction is dominant,~\cite{kugel} and can be explained by the present calculation where the exchange paths are properly taken into account in a realistic lattice. As for the no long-range SO orders in the hexagonal samples, the present SORS is a plausible candidate. The temperature dependence of the magnetic susceptibility is decomposed into the Curie tail and a gapped component. The former is attributed to the orphan spins, and the latter is explained by the present SORS where the short-range spin singlets are realized. 
Existence of the gapped magnetic excitation is also suggested by the specific heat, the inelastic neutron scattering and the nuclear magnetic resonance measurements.~\cite{Nakatsuji12,Quilliam12} The time scale for the SO dynamics in SORS is governed by the local DJT effect $J_{\rm DJT}\sim$1-10meV and the inter-site exchange interaction $J_{\rm SE} \sim $1-10meV, both of which are in between the ESR time scale ($\sim 10^{-9}$s) and the x-ray time scale ($\sim 10^{-15}$s). This fact can explain the contradicted experimental results: the almost isotropic ESR signal and the anisotropic extended x-ray absorption fine structure data,~\cite{Nakatsuji12} which are in contrast to the conventional strong JT coupling systems.~\cite{Sanchez03,Deisenhofer03}

Finally, our theory provides a number of forceful predictions in BSCO and other materials. There will be a crossover frequency/magnetic field in ESR, corresponding to $J_{\rm DJT}$ and $J_{\rm SE}$, where the anisotropy in the $g$-factor is changed qualitatively. Dynamics of the orbital-lattice coupled vibronic excitation is expected to be observed directly by inelastic light/x-ray scattering spectra around 1-10 meV. A key ingredient in the present SORS is the SO entanglement. To demonstrate the SO entanglement from the viewpoint of dynamics, we show, in Fig.~\ref{eta_dynamics}(b) and (c), the dynamical spin-correlation function $K_s(\omega)$ and the dynamical orbital-correlation function $K_\tau(\omega)$, respectively. 
We define 
\begin{align}
 K_u(\omega)=-\frac{1}{\pi}{\rm Im}\means{\mathcal{M}_u \frac{1}{\omega-({\cal H}+{\cal H}_{\rm inter})+E_0+i\eta}\mathcal{M}_u},
\end{align}
where $\mathcal{M}_u=\mathcal{M}_s$ for spin $(u=s)$, $\mathcal{M}_u=\frac{1}{6}\sum_i T_i^z$ for orbital $(u=\tau)$, an infinitesimal constant $\eta$, and the ground-state energy $E_0$. Gapped spin excitations and low-lying orbital excitations are seen in SORS and are consistent with the inelastic neutron and x-ray scattering experiments, respectively.~\cite{Nakatsuji12,Ishiguro2013} In SORS ($j_D=0.5$), in contrast to the orbital-ordered phase ($j_D=0.01$), an intensive orbital excitation is seen around the lowest spin-excitation energy, where we obtained that the SO correlation function $G$ is much larger than that in the ground state. This SO entangled excitation will be confirmed by combined analyses of the inelastic neutron and x-ray scattering experiments.
We also predict that the SORS is suppressed by applying the strong uni-axial pressure which breaks an energy balance between spin and orbital. Finally, in addition to Ba$_3$CuSb$_2$O$_9$, the present SORS scenario is applicable to other materials, where octahedra are not shared and JT centers are separated with each other. One plausible candidate is orbital-degenerate magnets in the ordered double-perovskite crystal lattice. 

In summary, we find that the SORS is realized by the DJT effect in a honeycomb lattice SO model. The present study provides systematic explanations for the recent experiments in Ba$_3$CuSb$_2$O$_9$. With increasing DJT, the local orbital moments are reduced, and the long-range orbital-ordered state is transferred to the quantum resonant state at the quantum-critical point. This interplay between the local quantum state and the classical order is analogous to the well known quantum-critical phenomena in the Kondo-lattice model. This theory also proposes a new route to the QSL state in orbitally degenerate systems without geometrical frustration. 

\begin{acknowledgments}
We thank S.~Nakatsuji, H.~Sawa, M.~Hagiwara and Y.~Wakabayashi for helpful discussions. This work was supported by KAKENHI from MEXT and Tohoku University ``Evolution'' program. JN is supported by the global COE program ``Weaving Science Web beyond Particle-Matter Hierarchy'' of MEXT, Japan. Parts of the numerical calculations are performed in the supercomputing systems in ISSP,  the University of Tokyo. 
\end{acknowledgments}



\clearpage


\begin{center}
{\large \bf Supplemental Material: 
Dynamical Jahn-Teller Effect in Spin-Orbital Coupled System
}
\end{center}

In this Supplemental Material, detailed derivations of the superexchange Hamiltonian and the effective vibronic Hamiltonian are presented.

\section{Superexchange Hamiltonian}
In this section, a derivation and explicit forms of the superexchange (SE) Hamiltonian in Eq.~(1) in the main article are presented.  

\subsection{Derivation of the superexchange Hamiltonian}
\label{suppl:sec:superex}
\begin{figure}[b]
\begin{center}
\includegraphics[width=0.7\columnwidth,clip]{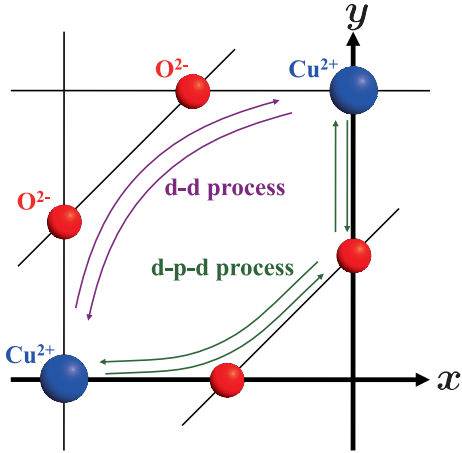}
\end{center}
\caption
{
Schematic two-types of the exchange processes on the $xy$ plane.
}
\label{suppl:sup_exch}
\end{figure}
The SE interaction Hamiltonian is derived from the extended $dp$-type Hamiltonian where Cu $3d$ and O $2p$ orbitals are considered, and the electron-electron interactions and the electron transfers are taken into account. All possible exchange paths between the NN Cu pairs are obtained by the perturbational procedures as follows.

Let us first consider a two-dimensional honeycomb lattice where a CuO$_6$ octahedron is located in each site as shown in Fig.~1(a) in the main article. An octahedron and a Cu site are labeled by an index $i$ and their positions are denoted by ${\bm r}_i$. The O sites which belong to the $i$-th octahedra are labeled by the indexes $i$ and $\delta(=\pm x, \pm y, \pm z)$. Positions of these O sites are ${\bm r}_i+{\bm d}_{\delta} $, where ${\bm d}_{\delta}$ is a vector along a direction $\delta$ with an amplitude of the nearest neighbor (NN) Cu-O bond distance. The axes of coordinates, $x, y, z$, are taken to be along the NN Cu-O bond directions in an octahedron as shown in Fig.~1(a) in the main article. A NN Cu-Cu bond, where the exchange paths are located on the $lm$-plane (see Fig.~\ref{suppl:sup_exch}), is labeled by an index $n$, where $(l, m, n)$ are the cyclic permutations of the Cartesian coordinates.  

We start from the extended $dp$-type Hamiltonian where the doubly-degenerate $e_g$ orbitals in Cu sites and the three $2p$ orbitals in O sites are introduced. The Hamiltonian is given by 
\begin{align}
{\cal H}_{pd}={\cal H}_{t}+{\cal H}_{\Delta}+{\cal H}_{Ud}+{\cal H}_{Up}.
\label{suppl:eq:hpd}
\end{align}
The first term represents the electron transfer between the Cu and O orbitals and that between the O orbitals:
\begin{align}
{\cal H}_t=
&-\sum_{i \delta}\sum_{\gamma \eta \sigma} \left( 
t_{dp; i \delta}^{\gamma \eta}
  d_{i \gamma \sigma }^\dagger p_{i \delta \eta \sigma}+H.c. \right)
  \nonumber \\
&-\sum_{<i\delta;j\delta'>}\sum_{\eta \eta' \sigma} \left(t_{pp; i \delta j \delta'}^{\eta\eta'}
  p_{i \delta \eta \sigma }^\dagger p_{j \delta' \eta' \sigma}
+H.c. \right),
\label{suppl:eq:ht}
\end{align}
where $d_{i\gamma \sigma}$ is the annihilation operator for a Cu hole at the $i$-th octahedron with orbital $\gamma(=3z^2-r^2,x^2-y^2)$ and spin $\sigma(=\uparrow, \downarrow)$, and $p_{i \delta \eta \sigma}$ is the annihilation operator for an O hole at the $\delta$-th O site in the $i$-th octahedron with orbital $\eta(=x, y, z)$ and spin $\sigma$. A symbol $<i\delta;j\delta'>$ represents the NN O sites labeled by $(i, \delta)$ and $(j, \delta')$, and  $ t_{dp; i \delta}^{\gamma \eta}$ and $t_{pp; i \delta j \delta'}^{\eta\eta'}$ are the corresponding transfer integrals. The second term in Eq.~(\ref{suppl:eq:hpd}) represents the energy differences given by  
\begin{align}
{\cal H}_{\Delta}
= \sum_{i \delta}\left[\left ( \Delta+\Delta_p \right) n_{i \delta \eta_\delta}^p
+\left ( \Delta-\Delta_p \right) \sum_{\eta \neq \eta_\delta} n_{i \delta \eta}^p
\right],\label{suppl:eq:2}
\end{align}
where $n_{i \delta \eta}^p (\equiv \sum_{\sigma}p_{i \delta \eta \sigma}^\dagger p_{i \delta \eta \sigma})$ is the number operator, and $\eta_\delta$ indicates the $2p_{|\delta|}$ orbital. The first (second) term represents the energy difference between the Cu $e_g$ and O $2p$ orbitals, which (do not) form the $\sigma$ bond, and $\Delta$ and $\Delta_p$ are the energy parameters.  
The third and fourth terms in Eq.~(\ref{suppl:eq:hpd}) represent the on-site electron-electron interactions in the Cu and O sites, respectively. These are given by 
\begin{align}
{\cal H}_{Ud}&=U_d\sum_{i\gamma}d_{i\gamma\uparrow}^\dagger d_{i\gamma\uparrow}
d_{i\gamma\downarrow}^\dagger d_{i\gamma\downarrow}
\nonumber \\
&+U'_d\sum_{i\sigma\sigma'}\sum_{\gamma>\gamma'}d_{i\gamma\sigma}^\dagger d_{i\gamma\sigma}
d_{i\gamma'\sigma'}^\dagger d_{i\gamma'\sigma'}
\nonumber \\
&-J_d\sum_{i\sigma\sigma'}\sum_{\gamma>\gamma'}d_{i\gamma\sigma}^\dagger d_{i\gamma\sigma'}
d_{i\gamma'\sigma'}^\dagger d_{i\gamma'\sigma} \nonumber \\
&-J'_d\sum_i \sum_{\gamma>\gamma'} \left (d_{i\gamma\uparrow}^\dagger d_{i\gamma\downarrow}
d_{i\gamma'\uparrow}^\dagger d_{i\gamma'\downarrow} +{\rm H.c.} \right ),
\label{suppl:eq:hud}
\end{align}
and 
\begin{align}
{\cal H}_{Up}&=
U_p\sum_{i\delta\eta}p_{i \delta \eta\uparrow}^\dagger p_{i\delta\eta\uparrow}
p_{i\delta\eta\downarrow}^\dagger p_{i\delta\eta\downarrow} \nonumber \\
&+U'_p\sum_{i \delta \sigma\sigma'}\sum_{\eta>\eta'}p_{i\delta\eta\sigma}^\dagger p_{i\delta\eta\sigma}
 p_{i\delta\eta'\sigma'}^\dagger p_{i\delta\eta'\sigma'}
\nonumber \\
& -J_p\sum_{i \delta \sigma\sigma'}\sum_{\eta>\eta'}p_{i\delta\eta\sigma}^\dagger p_{i\delta\eta\sigma'}
 p_{i\delta\eta'\sigma'}^\dagger p_{i\delta\eta'\sigma} \nonumber \\
& -J'_p\sum_{i\delta} \sum_{\eta>\eta'} \left(p_{i\delta\eta\uparrow}^\dagger p_{i\delta\eta\downarrow}
 p_{i\delta\eta'\uparrow}^\dagger p_{i\delta\eta'\downarrow} +{\rm H.c.} \right),
\label{suppl:eq:hup}
\end{align}
where $U_d$ ($U_p$), $U_d'$ ($U_p'$), $J_d$ ($J_p$) and $J_d'$ ($J_p'$) are the intra-orbital Coulomb interaction, the inter-orbital Coulomb interaction, the Hund coupling and the pair-hopping interaction for the $d$ ($p$) electrons, respectively. We assume the conditions $U_d=U_d'-2J_d$ and $J_d'=J_d$ in Eq.~(\ref{suppl:eq:hud}). 

From the $dp$-type Hamiltonian, the superexchange interactions between the NN Cu sites are obtained by the perturbational expansion. The exchange processes are classified into the two processes; two holes occupy virtually the same Cu (O) site in the intermediate states, termed the $dd$ ($dpd$) processes as shown in Fig.~\ref{suppl:sup_exch}. The superexchange interactions for the NN $ij$ sites labeled by $l$ through the $dd$ processes are given by 
\begin{align} 
{\cal H}_{dd}^{ij(l)}=&-A_d\left(
\frac{5}{4}-5\tau_i^{l}\tau_j^{l}
+3\bar{\tau}_i^{l}\bar{\tau}_j^{l}
+3T_i^y T_j^y\right)P_{ij}^T \nonumber\\
-B_d & \left(
\frac{5}{2}-2\tau_i^{l}-2\tau_j^{l}-6T_i^y T_j^y\right)P_{ij}^S\nonumber\\
-C_d& \left(
\frac{5}{4}-2\tau_i^{l}-2\tau_j^{l}
+5\tau_i^{l}\tau_j^{l}
-3\bar{\tau}_i^{l}\bar{\tau}_j^{l}
+3T_i^y T_j^y\right)P_{ij}^S,
\end{align}
where 
$A_d=t_p^2 t_d^4/[\Delta^4 (U'_d-J_d)]$, 
$B_d=t_p^2 t_d^4/[\Delta^4(U'_d+J_d)]$, and 
$C_d=t_p^2 t_d^4/[\Delta^4(U_d+J_d)]$.
The transfer integrals $t_d$ and $t_p$ are defined by the Slater-Koster parameters as $-t_d=(pd\sigma)$ and $-t_p=\frac{1}{2}(pp\sigma)+\frac{1}{2}(pp\pi)$, respectively.~\cite{Slater} The spin-singlet and spin-triplet operators are introduced by $P_{ij}^T=\frac{3}{4}+\bm{S}_i\cdot\bm{S}_j$ and $P_{ij}^S=\frac{1}{4}-\bm{S}_i\cdot\bm{S}_j$, respectively, where ${\bm S}_i$ is the spin operator with an amplitude of 1/2 at site $i$. The orbital degree of freedom in a Cu ion is represented by the pseudo-spin operator $\bm {T}_i$ with an amplitude of $1/2$. For convenience, we introduce $\tau_i^l=\cos(2 \pi n_l/3) T_i^z -\sin(2 \pi n_l/3) T_i^x$ and $\bar{\tau}_i^l=\cos(2 \pi n_l/3) T_i^x +\sin(2 \pi n_l/3) T_i^z$ with $(n_z, n_x, n_y)=(0,1,2)$. The eigen state of $\tau^l$ with the eigen value of +1/2 (-1/2) corresponds to the state where $d_{3l^2-r^2}$ $(d_{m^2-n^2})$ orbital is occupied by a hole.
In the same way, the exchange interactions through the $dpd$ processes are given by 
\begin{align}
 {\cal H}_{dpd}^{ij(l)}=&-A_p\left(
4-4\tau_i^{l}-4\tau_j^{l}
+4\tau_i^{l} \tau_j^{l}
-12\bar{\tau}_i^{l} \bar{\tau}_j^{l}
\right)P_{ij}^T\nonumber\\&
-(B_p+2C_p) \nonumber \\
& \times \left(
4-4\tau_i^{l}-4\tau_j^{l}
+4\tau_i^{l} \tau_j^{l}
-12\bar{\tau}_i^{l} \bar{\tau}_j^{l}
\right)P_{ij}^S,
\end{align}
where 
$A_p=t_p'^{2} t_d^4/ [\Delta^4(U'_p-J_p+2\Delta)]$,
$B_p=t_p'^{2} t_d^4/[\Delta^4 (U'_p+J_p+2\Delta)]$
and
$C_p=t_p^{2} t_d^4 (U_p+2\Delta-\Delta_p) /
[\Delta^4 \{ (U_p+2\Delta)^2-J_p'^2-\Delta_p^2 \}]$. 
The transfer integral $t'_p$ is defined by $-t'_p=\frac{1}{2}(pp\sigma)-\frac{1}{2}(pp\pi)$.~\cite{Slater} 

The two-types of the superexchange Hamiltonians are now taken together: 
\begin{align}
{\cal H}_{\rm exch}&=\sum_{<ij>_l} \left ({\cal H}_{dd}^{ij(l)}+{\cal H}_{dpd}^{ij(l)} \right)\nonumber\\
&=\sum_{<ij>_l}
\Bigl[J_{ss}\bm{S}_i\cdot\bm{S}_j+J_{\tau\tau}\tau_i^l \tau_j^l+J_{\bar{\tau}\bar{\tau}}\bar{\tau}_i^l \bar{\tau}_j^l+J_{yy}T_i^y T_j^y \nonumber \\
&+J_{ss\tau}\bm{S}_i\cdot\bm{S}_j (\tau_i^l +\tau_j^l)
+J_{ss\tau\tau}\bm{S}_i\cdot\bm{S}_j \tau_i^l \tau_j^l
\nonumber \\
&+J_{ss\bar{\tau}\bar{\tau}}\bm{S}_i\cdot\bm{S}_j \bar{\tau}_i^l \bar{\tau}_j^l+J_{ssyy}\bm{S}_i\cdot\bm{S}_j T_i^y T_j^y\Bigr], 
\label{suppl:eq:1}
\end{align}
where the exchange parameters are defined by 
$J_{ss}=-\frac{5}{4}A_d - 4 A_p + \frac{5}{2} B_d + 4 B_p +\frac{5}{4} C_d + 8 C_p$,
$J_{\tau\tau}=\frac{15}{4} A_d -3 A_p -  B_p -\frac{5}{4}  C_d - 2 C_p$, 
$J_{\bar{\tau}\bar{\tau}}=-\frac{9}{4} A_d + 9 A_p + 3 B_p +\frac{3}{4} C_d + 6 C_p$,
$J_{yy}=-\frac{9}{4}  A_d + \frac{3}{2}  B_d - \frac{3}{4} C_d$, 
$J_{ss\tau}=4 A_p -2 B_d -4 B_p -2 C_d - 8 C_p$, 
$J_{ss\tau\tau}=5 A_d - 4 A_p + 4 B_p + 5C_d + 8 C_p$, 
$J_{ss\bar{\tau}\bar{\tau}}=-3A_d +12 A_p -12 B_p -3 C_d -24 C_p$ and
$J_{ssyy}=-3A_d -6 B_d +3 C_d$.

We estimate the exchange constants quantitatively. By using the realistic parameter values,~\cite{Mizuno98} $t_{d}=1.2$eV, $t_{p}=t_{p}'=-0.65$eV, $U_d'=6$eV, $U_p=4$eV, $U_p'=3$eV, $J_d/U_d'=J_p/U_p'=0.5$, $\Delta=3$eV and $\Delta_p=0.1$eV, we have $J_{\tau\tau}/J_{ss}=0.78$, $J_{ss\tau\tau}/J_{ss}=4.5$, $J_{ss\tau}/J_{ss}=1.7$, $J_{\bar{\tau}\bar{\tau}}/J_{ss}=2.3$, $J_{ss\bar{\tau}\bar{\tau}}/J_{ss}=-5.2$, $J_{yy}/J_{ss}=-0.89$, and $J_{ssyy}/J_{ss}=-3.1$ as a unit of $J_{ss}=6.5$meV. These values are adopted to analyze the spin and orbital states in the two NN Cu sites. In the numerical calculations for ${\cal H}_{\rm eff}={\cal H}_{\rm exch}+{\cal H}_{\rm JT}$ in the main article, we adopt the above ratios of the exchange constants as a unit of $J_{ss}/J_{\rm AH}=0.15$.

\subsection{Generalization of the superexchange interaction}

In order to examine the connection between the superexchange Hamiltonian in Eq.~(\ref{suppl:eq:1}) and the honeycomb-lattice orbital-only model given by 
\begin{align}
{\cal H}_{\rm orb}=J\sum_{\langle ij \rangle_l} \tau_i^l \tau_j^l , 
\label{suppl:eq:only} 
\end{align}
we generalize the electron transfer integral and introduce the staggered magnetic field, as follows.

First, we generalize the transfer integral between the $p_x$ and $d_{3z^2-r^2}$ orbitals and that between $p_x$ and $d_{x^2-y^2}$ along the $x$ direction as $(t_{pd;i x}^{3z^2-r^2, x}, t_{pd; i x}^{x^2-y^2, x})=(-t_d/2, t_d\sqrt{3}/2) \rightarrow (t_{pd; i x}^{3z^2-r^2, x}(\eta), t_{pd; i x}^{x^2-y^2, x}(\eta))=(-t_d\cos\eta, t_d\sin\eta)$ by introducing a parameter $\eta$. The transfer integrals between other orbitals and those along other directions are obtained by the symmetry considerations. The generalized transfer integrals at $\eta=\pi/3$ reproduce the original transfer integrals. At $\eta=\pi/2$, we have $[t_{pd; i x}^{3z^2-r^2, x}(\eta=\pi/2), t_{pd; i x}^{x^2-y^2, x}(\eta=\pi/2)]=(0, t_d)$, i.e. the electron transfer between the $p_x$ and $d_{3z^2-r^2}$ orbitals along $z$ vanishes. 

Through the generalization of the transfer integrals, we derive the superexchange Hamiltonian. The generalized exchange constants are given by 
$J_{ss}(\eta)=-A_d - 4 A_p + 2 B_d + 4 B_p + C_d +  8 C_p + (-A_d + 2 B_d + C_d) \cos^2 2\eta$, 
$J_{\tau\tau}(\eta)=3 A_d - C_d + (3 A_d - 12 A_p - 4 B_p - C_d - 8 C_p) \cos^2 2\eta$,
$J_{\bar{\tau}\bar{\tau}}(\eta)=(-3 A_d + 12 A_p + 4 B_p + C_d + 8 C_p) \sin^22\eta$, 
$J_{yy}(\eta)=-(3 A_d - 2 B_d + C_d) \sin^2 2\eta$, 
$J_{ss\tau}(\eta)=4 (2 A_p - B_d - 2 B_p - C_d - 4 C_p) \cos 2\eta$, 
$J_{ss\tau\tau}(\eta)=4 (A_d + C_d + (A_d - 4 A_p + 4 B_p + C_d + 8 C_p) \cos^2 2\eta)$, 
$J_{ss\bar{\tau}\bar{\tau}}(\eta)=-4 (A_d - 4 A_p + 4 B_p + C_d + 8 C_p) \sin^2 2\eta$, 
and 
$J_{ssyy}(\eta)=-4 (A_d + 2 B_d - C_d) \sin^2 2\eta$. At $\eta=\pi/2$, the superexchange Hamiltonian is given by 
\begin{align}
{\cal H}_{\rm exch}&=
\sum_{<ij>_l}
\bigl [
J_{ss}\bm{S}_i\cdot\bm{S}_j+J_{\tau\tau}\tau_i^l \tau_j^l \nonumber \\
&+J_{ss\tau}\bm{S}_i\cdot\bm{S}_j (\tau_i^l +\tau_j^l)
+J_{ss\tau\tau}\bm{S}_i\cdot\bm{S}_j \tau_i^l \tau_j^l
\bigr ] . 
\label{suppl:eqpi/2}
\end{align}
To reproduce the orbital-only model, we further introduce the staggered magnetic field defined by 
\begin{align}
 \mathcal{H}_h=-h \sum_i(-1)^iS_i^z . 
\end{align}
In the limit of $h \rightarrow \infty $, the spin degree of freedom is frozen, and Eq.~(\ref{suppl:eqpi/2}) is reduced to the orbital-only model in Eq.~(\ref{suppl:eq:only}). The exchange constant $J$ corresponds to $J_{\tau\tau}(\eta=\pi/2)-J_{ss\tau\tau}(\eta=\pi/2)/4$ which is negative.

\section{vibronic Hamiltonian}
\label{suppl:sec:deriv-low-energy}
In this section, a detailed derivation of the effective vibronic Hamiltonian in Eq.~(5) in the main article is presented.  

\subsection{Derivation of the effective vibronic Hamiltonian}
\label{suppl:sec:vib}
\begin{figure}[b]
\begin{center}
\includegraphics[width=0.9\columnwidth,clip]{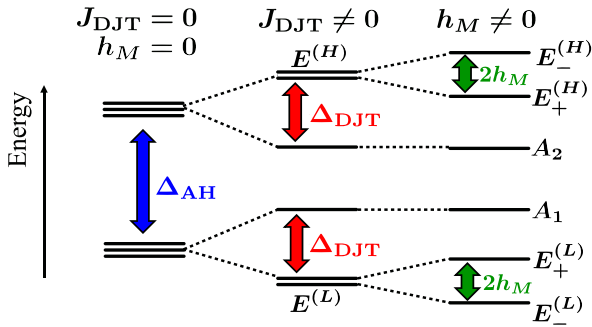}
\end{center}
\caption
{
The energy level scheme for the vibronic states.
}
\label{suppl:sup_energy}
\end{figure}

We start from the electron-lattice interaction Hamiltonian. The two vibrational modes, $Q_u$ and $Q_v$, with the $E$ symmetry couple with the $e_g$ electronic orbitals. This is given by 
\begin{align}
 {\cal H}_{e\textrm{-}l}&=\sum_i\Biggl[
-\frac{1}{2M}\left(\frac{\partial^2}{\partial Q_{ui}^2}+\frac{\partial^2}{\partial Q_{vi}^2}\right)
+\frac{M\omega^2}{2}\rho_i^2
\nonumber\\ 
&+2A(T_i^x Q_{vi}+T_i^z Q_{ui})+B(Q_{ui}^3-3Q_{vi}^2 Q_{ui})\Biggr] , 
\label{suppl:eq:el}
\end{align}
where $\rho_i=\sqrt{Q_{ui}^2+Q_{vi}^2}$ is an amplitude of the lattice distortion at the $i$-th octahedron. The first two terms are for the harmonic vibrations with frequency $\omega$ and oxygen mass $M$. The third term describes the Jahn-Teller (JT) coupling with a coupling constant $A(>0)$, and the last term represents the anharmonic lattice potential, where $B$ is negative. 

We analyze this Hamiltonian based on the Born-Oppenheimer approximation. The vibronic wave-function is written as a product of the electronic wave function, $\psi_k$, and the lattice wave function, $\phi_\lambda^k$, as $\Phi_{\lambda,k}(\bm{r},\bm{Q})=\psi_k(\bm{r},\bm{Q})\phi_\lambda^k(\bm{Q})$, where $\bm{r}$ and $\bm{Q}$ are  the electron and lattice coordinates, respectively, and $k$ and $\lambda$ describe the electronic and vibrational states, respectively. The adiabatic potentials for an octahedron are given as 
\begin{align}
 U^{(k=\pm)}=\frac{M\omega^2\rho^2}{2}\pm A\rho + B\rho^3 \cos\theta,
\end{align}
where $\theta \equiv \tan^{-1}(Q_v/Q_u)$. The wave functions for a hole corresponding to the lower adiabatic potential, i.e. $k=-$, is given as 
\begin{align}
\psi_{-}(\bm{r},\theta)=\psi_{x^2-y^2}(\bm{r})\cos\frac{\theta}{2}-\psi_{3z^2-r^2}(\bm{r})\sin\frac{\theta}{2}.\label{suppl:eq:psi}
\end{align}
In the case of $B=0$, $U^{(-)}$ takes its minimum of $E_{\rm JT}=A^2/(2M\omega^2)$ at $\rho=\rho_0\equiv A/(M\omega^2)$ for any $\theta$. When $B$ is taken into account, the potential takes its minima (maxima) at angles $\theta_{\mu \nu}=\mu\pi+2\nu\pi/3$ with integers $\mu=0$ ($\mu=1$) and $\nu=(0,1,2)$, as shown in Fig.~1(b) in the main article. 

Here we assume that the zero-point vibration energy ($\omega /2 $) is sufficiently smaller than the JT energy ($E_{\rm JT}$), and the vibronic motion is confined on the lower adiabatic potential. Then, the effective Hamiltonian is given by 
\begin{align}
 {\cal H}_{\rm vib} &=\int d\bm{r}\psi_{-}(\bm{r},\bm{Q})^*
\bigl[
-\frac{1}{2M}\left(
\frac{\partial^2}{\partial Q_u^2}+\frac{\partial^2}{\partial Q_v^2}
\right)
+U^{(-)}
\bigr]
\nonumber \\
& \ \ \ \ \ \ \ \times \psi_{-}(\bm{r},\bm{Q}) \nonumber\\
&=\frac{1}{\sqrt{\rho}}\left[
-\frac{1}{2M}\frac{\partial^2}{\partial\rho^2}
+\frac{M\omega^2}{2}(\rho-\rho_0)^2-E_{\rm JT}
\right]\sqrt{\rho} \nonumber \\
&-\frac{1}{2M \rho^2}
\frac{\partial^2}{\partial \theta^2} +B \rho^3 \cos 3\theta.\label{suppl:eq:7}
\end{align}
The vibronic motion is classified into the radial mode where the amplitude $\rho$ varies around $\rho_0$, and the rotational mode where the angle $\theta$ varies at $\rho=\rho_0$. We consider the case that the excitation energy for the radial mode, $\omega$, is larger than the kinetic energy  for rotational mode, $1/(2M \rho_0^2)=\omega^2/(4E_{\rm JT})$, and focus on the rotational mode, corresponding to the last two terms in Eq.~(\ref{suppl:eq:7}). The effective Hamiltonian for the rotational mode is defined as 
\begin{align}
{\cal H}_{\rm rot}=-\frac{1}{2M \rho_0^2}
\frac{\partial^2}{\partial \theta^2} +B \rho_0^3 \cos 3\theta . 
\label{suppl:eq:3}
\end{align}
Since this Hamiltonian describes the kinetic motion under the periodic potential, solutions of the Schr\"odinger equation, ${\cal H}_{\rm rot}\phi_\lambda(\theta)=\varepsilon_\lambda \phi_\lambda (\theta)$, are obtained as the Bloch states.~\cite{Koehler40,OBrien64,Williams69} As shown in Fig.~\ref{suppl:sup_energy}, the lowest-six eigen states are labeled by the irreducible representation in the C$_{\rm 3v}$ group, $E^{(L)}$, $A_1$, $A_2$ and $E^{(H)}$, where $E^{(L)}$ and $E^{(H)}$ are the doubly degenerate representations and the degenerate bases are labeled by the indexes $+$ and $-$. From the Bloch-type eigen states, we introduce the Wannier-type wave functions. From the lowest-six eigen states, we introduce the six Wannier functions,  $\phi_{0 \nu}(\theta)$ and $\phi_{1 \nu}(\theta)$ $(\nu=0, 1, 2)$, which are almost localized around $\theta_{0 \nu}$ and $\theta_{1 \nu}$ respectively. Explicit relations between the Bloch functions and the Wannier functions are given by 
$\phi_{E^{(L)}_\pm}=(\phi_{00}+e^{\pm 2i\pi/3}\phi_{01}+e^{\pm 4i\pi/3}\phi_{02})/\sqrt{3}$, 
$\phi_{A_1}=(\phi_{00}+\phi_{01}+\phi_{02})/\sqrt{3}$, 
$\phi_{A_2}=(\phi_{10}+\phi_{11}+\phi_{12})/\sqrt{3}$ and 
$\phi_{E^{(H)}_\pm}=(\phi_{10}+e^{\pm 2i\pi/3}\phi_{11}+e^{\pm 4i\pi/3}\phi_{12})/\sqrt{3}$.
Because of the parity of the wave functions, the two spaces based on the wave functions $\phi_{0 \nu}(\theta)$ and  $\phi_{1 \nu}(\theta)$, termed the $\mu$=0 and 1 spaces, respectively, are orthogonal with each other and are treated independently.

For simplicity, we assume that the energy difference between the $E^{(L)}$ and $A_1$ states is equal to that between $A_2$ and $E^{(H)}$. We denote this energy difference by $\Delta_{\rm DJT}$ and that between the $A_1$ and $A_2$ states by  $\Delta_{\rm AH}$. Then, the low-energy vibronic Hamiltonian is obtained as a simple form: 
\begin{align}
{\cal H}_{\rm JT}
&=\left( \Delta_{\rm DJT}+\frac{\Delta_{\rm AH}}{2} \right)
 \nonumber \\& \times
 \left( 
 -\kets{\Phi_{E^{(L)}_+}}\bras{\Phi_{E^{(L)}_+}}  
-\kets{\Phi_{E^{(L)}_-}}\bras{\Phi_{E^{(L)}_-}}  \right.\nonumber \\
&\ \ \ \ \  \ \left.
 \kets{\Phi_{E^{(H)}_+}}\bras{\Phi_{E^{(H)}_+}} 
+\kets{\Phi_{E^{(H)}_-}}\bras{\Phi_{E^{(H)}_-}} \right ) \nonumber \\
&+\frac{\Delta_{\rm AH}}{2}
\left ( 
-\kets{\Phi_{A_1}}\bras{\Phi_{A_1}}
+\kets{\Phi_{A_2}}\bras{\Phi_{A_2}} \right ). 
\end{align}
This is rewritten in the Wannier function representation as 
\begin{align}
{\cal H}_{\rm JT}=
  \sum_{i \mu}\frac{\sigma_\mu}{2}\Bigl[-J_{\rm AH}\sum_{\nu}\kets{\Phi_{i \mu\nu}}\bras{\Phi_{i \mu\nu}}\nonumber\\
+J_{\rm DJT}\sum_{\nu\neq \nu' }\kets{\Phi_{i \mu\nu}}\bras{\Phi_{i \mu\nu'}}\Bigr],
\label{suppl:eq:djt}
\end{align}
where $(\sigma_0, \sigma_1)=(1,-1)$ for the index $\mu$. We define $J_{\rm DJT}=2\Delta_{\rm DJT}/3$ and $J_{\rm AH}=4\Delta_{\rm DJT}/3+\Delta_{\rm AH}$, by which the condition $J_{\rm DJT}/J_{\rm AH}<1/2$ is derived. 

On an equal footing of the effective vibronic Hamiltonian in Eq.~(\ref{suppl:eq:djt}), we provide a representation for the superexchange Hamiltonian, where the six vibronic wave functions, $\Phi_{\mu \nu}$ $(\mu=0, 1; \nu=0, 1,2)$, are adopted as a basis set. This is performed by introducing the projection operator as ${\cal H}_{\rm exch} \rightarrow {\cal P} {\cal H}_{\rm exch} {\cal P}$. Here ${\cal P}$ operates the vibronic wave function so as to be restricted within the six basis functions $\Phi_{\mu \nu}$ by changing a value of $\theta$. 

Finally, we show the Ham's reduction effect~\cite{Ham65,Ham68} in the present formulation. The vibronic wave-function is represented by a product form of $\Phi_{\mu \nu}({\bm r}; \theta) = \psi_-({\bm r}; \theta) \phi_{\mu \nu}(\theta)$. 
We introduce an approximation as 
$\Phi_{\mu \nu}(\bm{r};\theta)= \psi_-(\bm{r},\theta)\phi_{\mu \nu}(\theta)
\sim\psi_-(\bm{r},\theta_{\mu \nu})\phi_{\mu \nu}(\theta)$, 
where $\phi_{\mu \nu}(\theta)$ takes its maximum at $\theta_{\mu \nu}$. 
The lowest two eigen states of the Hamiltonian are given by 
\begin{align}
 \kets{\Phi_{E^{(L)}u}}&=\frac{1}{\sqrt{6}}(2\kets{\Phi_{00}}-\kets{\Phi_{01}}-\kets{\Phi_{02}}) , \\
 \kets{\Phi_{E^{(L)}v}}&=\frac{1}{\sqrt{2}}(\kets{\Phi_{01}}-\kets{\Phi_{02}}) , 
\end{align}
where
$\kets{\Phi_{00}}=\kets{\psi_{x^2-y^2}}\kets{\phi_{00}}$,  
$\kets{\Phi_{01}}=[-(1/2)\kets{\psi_{x^2-y^2}}-(\sqrt{3}/2)\kets{\psi_{3z^2-r^2}}]\kets{\phi_{01}}$, 
and 
$\kets{\Phi_{02}}=[-(1/2)\kets{\psi_{x^2-y^2}}+(\sqrt{3}/2)\kets{\psi_{3z^2-r^2}}]\kets{\phi_{02}}$. 
By using these wave functions, we obtain the matrix elements for the orbital pseudo-spin operators as 
\begin{align}
 T^z=\frac{1}{4}\bordermatrix{
& \Phi_{E^{(L)}u} & \Phi_{E^{(L)}v} \cr
 & 1 & 0 \cr
 & 0 & -1 \cr},
 \end{align}
and 
\begin{align}
T^x=\frac{1}{4}\bordermatrix{
& \Phi_{E^{(L)}u} & \Phi_{E^{(L)}v} \cr
 & 0 & 1 \cr
 & 1 & 0 \cr} ,
\end{align}
which imply that the pseudo-spin moment is reduced due to the quantum mechanical superposition. 
The Ham's reduction factor is obtained as $q \equiv 2\bras{\Phi_{E^{(L)}u}} T^z \kets{\Phi_{E^{(L)}u}}=-2\bras{\Phi_{E^{(L)}v}} T^z \kets{\Phi_{E^{(L)}v}}=2\bras{\Phi_{E^{(L)}u}} T^x \kets{\Phi_{E^{(L)}v}}=1/2$. 
This value is close to the previous results obtained by the strong coupling approximations.~\cite{Ham68, Williams69,Slonczewski69,Englman_text,Bersuker_text,Halperin69}

\subsection{Artificial field for vibronic state}
In the main article, we examine a connection between the  Hamiltonian ${\cal H}_{\rm eff}={\cal H}_{\rm exch}+{\cal H}_{\rm JT}$ and the Heisenberg model ${\cal H}_{\rm spin}=J\sum_{\langle ij \rangle} \bm{S}_i \cdot \bm{S}_j$ by introducing the artificial field for the orbital-lattice sector. We introduce the external field which acts on the eigen states $\Phi_{E^{(L)}_\pm}$ and $\Phi_{E^{(H)}_\pm}$ as 
\begin{align}
\mathcal{H}_M=-h_M \biggl (&
 \kets{\Phi_{E^{(L)}_-}}\bras{\Phi_{E^{(L)}_-}}
-\kets{\Phi_{E^{(L)}_+}}\bras{\Phi_{E^{(L)}_+}} \nonumber \\
-&\kets{\Phi_{E^{(H)}_-}}\bras{\Phi_{E^{(H)}_-}}
 +\kets{\Phi_{E^{(H)}_+}}\bras{\Phi_{E^{(H)}_+}} \biggr ) . 
\label{suppl:eq:hm}
\end{align}
In the Wannier function representation, we have 
\begin{align}
\mathcal{H}_M=-h_M \sum_{i\mu} \sigma_\mu {\cal F}_{i\mu}.
\end{align}
The $3 \times 3$ matrix is defined by $({\cal F}_\mu)_{\nu\nu'}=F_{\nu\nu'}|  \Phi_{i \mu \nu} \rangle \langle  \Phi_{i \mu \nu'}|$ for the index $\nu$, where $F_{\nu \nu'}=\frac{i}{\sqrt{3}}\sum_l\varepsilon_{l\nu \nu'}$ with the Levi-Civita completely-antisymmetric tensor $\varepsilon_{l\nu\nu'}$.
 The matrix ${\cal F}_\mu$ is explicitly given as 
\begin{align}
 {\cal F}_{i\mu}=\frac{1}{\sqrt{3}}
\bordermatrix{
& \Phi_{\mu 0} & \Phi_{\mu 1} & \Phi_{\mu 2} \cr
 & 0 & i & -i \cr
 & -i & 0 & i \cr
 & i & -i & 0 \cr }.
\end{align}
This is equivalent to the spin-orbit interaction Hamiltonian in the previous articles.~\cite{Williams69}

A schematic energy levels under the external field is given in Fig.~\ref{suppl:sup_energy}. Double degeneracies in the $E^{(L)}$ and $E^{(H)}$ levels are lifted by applying the virtual field. In the limit of $h_M \rightarrow \infty$, the orbital and vibrational degrees of freedom are frozen, and ${\cal H}_{\rm eff}={\cal H}_{\rm exch}+{\cal H}_{\rm DJT}$ is reduced into the Heisenberg model ${\cal H}_{\rm spin}$ on a honeycomb lattice. 

\subsection{Interaction between the NN O$_6$ octahedra}
\begin{figure}[b]
\begin{center}
\includegraphics[width=0.9\columnwidth,clip]{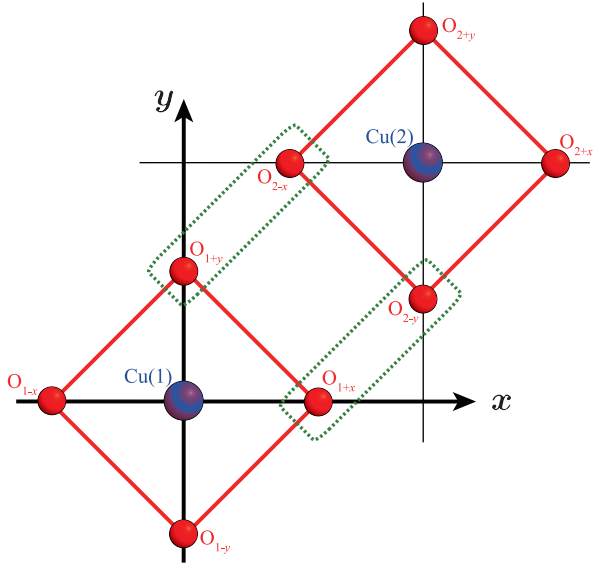}
\end{center}
\caption
{
The NN O$_6$ octahedra in the $x$-$y$ plane. Elastic interactions between the two oxygen ions surrounded by dotted lines are considered.}
\label{suppl:interlatt}
\end{figure}

So far, the each O$_6$ octahedron in a two-dimensional layer is assumed to be independent with each other. This is a reasonable approximation as the first step, since the NN two octahedra do not share the O ions, unlike the perovskite lattice structure. 
In this subsection, we introduce the interaction between the NN two O$_6$ octahedra which modifies the spin-orbital phase diagram as shown in Fig.~3 in the main article. 

As shown in Fig.~\ref{suppl:interlatt}, we consider a NN pair of the octahedra termed 1 and 2, in which the connecting Cu-O-O-Cu bonds are in the $xy$ plane. We consider the elastic interactions between O$_{1+x}$ and O$_{2-y}$, and that between O$_{1+y}$ and O$_{2-x}$. This is given as 
\begin{align}
 {\cal H}_{\rm elastic}^{z}=\frac{k}{2}(\Delta x_{1+x}-\Delta y_{2-y})^2
+\frac{k}{2}(\Delta y_{1+y}-\Delta x_{2-x})^2,
\end{align}
where $\Delta \bm{r}_{i\delta}$ represents the displacement of the O$_{i\delta}$ ion from the position without the Jahn-Teller distortion, and $k$ is the spring constant. Then, the interaction terms between the NN pair of the octahedra are obtained as 
\begin{align}
 {\cal H}_{\rm inter}^{z}=-k(\Delta x_{1+x}\Delta y_{2-y}+ \Delta y_{1+y}\Delta x_{2-x}).
\label{suppl:eq:100}
\end{align}
This is represented by the normal vibration modes $Q_{iu}$ and $Q_{iv}$ for the octahedron defined by 
\begin{align}
 Q_{iu}&=\frac{1}{\sqrt{6}}(-\Delta x_{i+x}+\Delta x_{i-x}-\Delta y_{i+y}+\Delta y_{i-y}\nonumber\\
&\ \ \ \ \ \ \ +2\Delta z_{i+z}-2\Delta z_{i-z}) , \\
 Q_{iv}&=\frac{1}{\sqrt{2}}(\Delta x_{i+x}-\Delta x_{i-x}+\Delta y_{i+y}-\Delta y_{i-y}).
\end{align}
as 
\begin{align}
 {\cal H}_{\rm inter}^z=\frac{K}{4 \rho_0^2} (3Q_{iu}Q_{ju}-Q_{iv}Q_{jv}) , 
\end{align}
with the coupling constant $K=(2\rho_0)^2k$.
In general, the interaction Hamiltonian between the NN $i$ and $j$ Cu sites, where the connecting bonds are perpendicular to the $l$ axis, is given by 
\begin{align}
 {\cal H}_{\rm inter}=\frac{K}{4 \rho_0^2}\sum_{<ij>_l} (3Q_{iu}^lQ_{ju}^l-Q_{iv}^lQ_{jv}^l) , 
\end{align}
where $Q_{iu}^l=\cos (2 n_l \pi /3) Q_{iu} +\sin(2n_l \pi/3) Q_{iv}$ and $Q_{iv}^l=-\sin (2 n_l \pi /3) Q_{iu} +\cos(2n_l \pi/3) Q_{iv}$ with $(n_z,n_x, n_y)=(0,1,2)$.

\section{Details of numerical calculation methods}

In this section, we present details of the calculation methods, the exact diagonalization with the mean-field approximation (ED+MF) and the quantum Monte-Carlo method with the mean-field approximation (QMC+MF), which are used to analyze the model Hamiltonian in the main article. 

\begin{figure}[b]
\begin{center}
\includegraphics[width=0.9\columnwidth,clip]{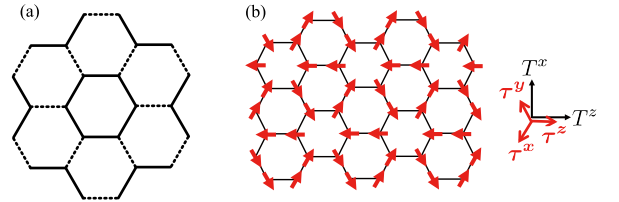}
\end{center}
\caption
{
(a) A honeycomb lattice structure adopted in the ED+MF method. Bold and dotted lines represent the bonds where the exchange interactions are treated exactly, and the bonds where the MF decouplings are applied, respectively. (b) The three-fold orbital ordered state introduced in the QMC+MF method. The red arrows represent the orbital PSs in the $T^z$-$T^x$ plane.
}
\label{suppl:assumptions}
\end{figure}

\subsection{Exact diagonalization with the mean-field approximation}

We consider the Hamiltonian ${\cal H}={\cal H}_{\rm exch}+{\cal H}_{\rm JT}$ in a honeycomb lattice. The lattice is divided into the hexagon clusters represented by the thick lines in Fig.~\ref{suppl:assumptions}(a) which are connected with each other by the dotted lines. We assume that all hexagon clusters are equivalent. The exchange Hamiltonian is treated exactly for the nearest-neighbor bonds in the hexagon clusters and approximately for connecting bonds between the clusters. We apply the mean-field decouplings to ${\cal H}_{\rm exch}$ for the connecting bonds as 
$\tau_i\tau_j\rightarrow \tau_i\means{\tau_j}+\means{\tau_i}\tau_j-\means{\tau_i}\means{\tau_j}$, $\bm{S}_i\cdot\bm{S}_j\rightarrow \bm{S}_i\cdot\means{\bm{S}_j}+\means{\bm{S}_i}\cdot\bm{S}_j-\means{\bm{S}_i}\cdot\means{\bm{S}_j}$ and $\bm{S}_i\cdot\bm{S}_j\tau_i\tau_j\rightarrow\bm{S}_i\tau_i\cdot\means{\bm{S}_j\tau_j}+\means{\bm{S}_i\tau_i}\cdot\bm{S}_j\tau_j-\means{\bm{S}_i\tau_i}\cdot\means{\bm{S}_j\tau_j}$. 
The first two are the conventional MF decouplings, and the last one implies that the intra-site spin and orbital correlations are taken into account exactly, but the inter-site ones are approximately. Since ${\cal H}_{\rm JT}$ describes the on-site interactions, the Hamiltonian in a hexagon under the mean-fields is analyzed by the exact-diagonalization method based on the Lanczos algorithm. 
In the numerical calculations, the 16,777,216 dimensions of the Hilbert space are reduced into 933,120 by utilizing the symmetry due to the $z$-component of the total-spin quantum number. By using the obtained ground-state wave-function, the expectation values, $\means{\tau_i}$, $\means{\bm{S}_i}$ and $\means{\tau_i\bm{S}_i}$, are calculated at each site, and the mean-fields acting on the cluster are obtained. These procedures are repeated until all expectation values converge. In the whole parameter regions shown in the main article, we confirm that the ground state is not degenerate. 

Since a hexagon is the smallest cluster, in which the expected spin/orbital fluctuations and orders are realized, the present ED+MF calculation is a minimal method to examine competition among the possible spin-orbital states. We have applied this kind of ED+MF method to other frustrated spin and orbital models, the so-called $J_1$-$J_2$ model where the NN and 2nd NN exchange interactions exist, and the ring-exchange model, and confirmed reliability of this method.~\cite{nasu11,Isaev2009,Lauchli05}

\subsection{Quantum Monte-Carlo method with the mean-field approximation}

This method is utilized to analyze spin structures in the orbital ordered states in the case of small DJT effect. We start from the Hamiltonian ${\cal H}={\cal H}_{\rm exch}+{\cal H}_{e\textrm{-}l}$ where the first and second terms are given by Eqs.~(\ref{suppl:eq:1}) and (\ref{suppl:eq:el}), respectively. From this Hamiltonian, we derive the effective spin Hamiltonian through the following procedures. 

First, we apply the mean-field approximation to the exchange Hamiltonian ${\cal H}_{\rm exch}$. 
The interactions between the orbital PSs, and those between the spins and orbital PSs are decoupled as 
$\tau_i\tau_j\rightarrow \tau_i\means{\tau_j}+\means{\tau_i}\tau_j-\means{\tau_i}\means{\tau_j}$,  $\tau_i\tau_j\bm{S}_i\cdot\bm{S}_j\rightarrow \tau_i\means{\tau_j}\means{\bm{S}_i\cdot\bm{S}_j}+\means{\tau_i}\tau_j\means{\bm{S}_i\cdot\bm{S}_j}+\means{\tau_i}\means{\tau_j}\bm{S}_i\cdot\bm{S}_j-2\means{\tau_i}\means{\tau_j}\means{\bm{S}_i\cdot\bm{S}_j}$ and so on.
Then, the Hamiltonian is given by 
\begin{align}
{\cal H}=
\sum_{<ij>} J_{ij} \bm{S}_i\cdot\bm{S}_j
+\sum_i {\cal H}_{i} ,
\label{suppl2:eq:5}
\end{align}
with 
\begin{align}
{\cal H}_{i}= - {\bm h}^{\rm MF}_i \cdot {\bm T}_i+{\cal H}_{e\textrm{-}l i } ,
\label{suppl2:eq:5b}
\end{align}
where $J_{ij}$ is the effective exchange interactions, ${\bm h}_{i}^{\rm MF}$ is the mean-field for the orbital PS operator at site $i$, and ${\cal H}_{e\textrm{-}l i }$ is the $i$-site term in ${\cal H}_{e\textrm{-}l } $. Here, we assume the three-fold orbital ordered state shown in Fig.~\ref{suppl:assumptions}(b) as a ground-state orbital structure in the case of a small DJT effect. This is suggested by the additional analyses of ${\cal H}_{\rm exch}$ by utilizing the ED method as well as the mean-field method. Under this orbital ordered state, a direction of ${\bm h}_{i}^{\rm MF}$ is parallel to $ {\bm T}_i$. Amplitudes of the PS operators at all sites are equal with each other and are given by $m=|\means{\bm T}|$. The exchange interactions, $J_{ij}$, in the first term in Eq.~(\ref{suppl2:eq:5}) are classified into the following two constants, 
\begin{align}
 J_1 = J_{ss}+2J_{ss\tau} m+J_{ss\tau\tau} m^2 , 
\label{suppl2:eq:1}
\end{align}
for the NN bonds where the PS operators are parallel with each other, and 
\begin{align}
 J_2 = J_{ss}-J_{ss\tau} m+\frac{1}{4}J_{ss\tau\tau} m^2 -\frac{3}{4}J_{ss\bar{\tau}\bar{\tau}}m^2 , \label{suppl2:eq:2}
\end{align}
for other NN bonds. As for the first term in Eq.~(\ref{suppl2:eq:5b}), amplitude of the mean-field is given by 
\begin{align}
h^{\rm MF} =&-\frac{3}{2}J_{\tau\tau}m + \frac{3}{2} J_{\bar{\tau}\bar{\tau}}m+ \frac{3}{2}J_{ss\bar{\tau}\bar{\tau}}\means{\bm{S}_i\cdot\bm{S}_j}_2 m\nonumber\\ &-J_{ss\tau\tau}(\means{\bm{S}_i\cdot\bm{S}_j}_1+\means{\bm{S}_i\cdot\bm{S}_j}_2/2)m\nonumber\\
& - J_{ss\tau}(\means{\bm{S}_i\cdot\bm{S}_j}_1-\means{\bm{S}_i\cdot\bm{S}_j}_2),\label{suppl2:eq:3}
\end{align}
where 
$\means{\bm{S}_i\cdot\bm{S}_j}_1$ and $\means{\bm{S}_i\cdot\bm{S}_j}_2$ are the spin correlations for the bonds where the effective exchange constants are $J_1$ and $J_2$, respectively. In the case of $h_{\rm MF}\ll E_{\rm JT}=A^2/(2M\omega^2)$, the second term in Eq.~(\ref{suppl2:eq:5b}) is reduced into the effective vibronic Hamiltonian which is similar to the model shown in Eq.~(\ref{suppl:eq:3}). Then, the low-energy effective model for Eq.~(\ref{suppl2:eq:5b}) is given by   
\begin{align}
{\cal H}_{i}= -\frac{1}{2M \rho_0^2}
\frac{\partial^2}{\partial \theta_i^2} +B \rho_0^3 \cos 3\theta_i  +\frac{1}{2}h^{\rm MF}\cos\theta_i. 
\label{suppl2:eq:rot2}
\end{align}

The Hamiltonian ${\cal H}$ in Eq.~(\ref{suppl2:eq:5}) with Eq.~(\ref{suppl2:eq:rot2}) is solved self-consistently, as follows. Under a given $m$, the first term in Eq.~(\ref{suppl2:eq:5}) is analyzed by utilizing the QMC method. The continuous imaginary-time method with the loop algorithm in the ALPS library is applied.~\cite{ALPS,Todo} Cluster size is chosen to be $20\times 20 \times 6$ and temperature is chosen to be $0.01J_1$. By using the calculated spin correlation functions $\means{\bm{S}_i\cdot\bm{S}_j}_1$ and $\means{\bm{S}_i\cdot\bm{S}_j}_2$, the mean-field acting on the orbital PS operator, $h^{\rm MF}$, is obtained. Then, we solve the Shr\"odinger equation ${\cal H}_{i}\phi_\mu(\theta)=\varepsilon_\mu\phi_\mu(\theta)$. The vibronic wave-function for the ground state is represented by $\Phi_0(\bm{r},\theta)=\psi_-(\bm{r};\theta)\phi_0(\theta)$ where $\phi_0(\theta)$ is the ground-state wave function for this Shr\"odinger equation and $\psi_-(\bm{r};\theta)$ is the electronic wave-function defined in Eq.~(\ref{suppl:eq:psi}). Finally, we have $\means{{\bm T}}=\bras{\Phi_{0}} {\bm T} \kets{\Phi_{0}}$. This procedure is repeated until $|\means{\bm T}|$ converges.


\begin{thebibliography}{99}
\bibitem{coord}
The axes of coordinate, $x, y, z$, are chosen to be parallel to the orthogonal Cu-O bond directions in an octahedron. Since the exchange paths are identified by a plane $(lm)$, a NN Cu-Cu bond is labeled by an index $n$, where $(l, m, n)$ are the cyclic permutations of the Cartesian coordinates.
\end{thebibliography}
\end{document}